\documentstyle[aps,prb,multicol,psfig]{revtex}

\begin{document}

\title{Temperature dependence of polarization relaxation in 
       semiconductor quantum dots}

\author{E.~Tsitsishvili\cite{ad} and R.~v.~Baltz}
\address{Institut f\"ur Theorie der Kondensierten Materie, 
Universit\"at Karlsruhe, D-76128 Karlsruhe, Germany}
\author{H.~Kalt}
\address{Institut f\"ur Angewandte Physik, 
Universit\"at Karlsruhe,  D-76128 Karlsruhe, Germany}

\maketitle

\begin{abstract}
The decay time of the linear polarization degree of the luminescence in 
strongly confined semiconductor quantum dots with asymmetrical shape is 
calculated in the frame of second-order quasielastic interaction between 
quantum dot charge carriers and LO phonons. The phonon bottleneck
does not prevent significantly the relaxation processes and the 
calculated decay times can be of the order of a few tens picoseconds at 
temperature  $T \simeq 100$K, consistent with recent experiments by 
Paillard et al. [Phys. Rev. Lett. {\bf86}, 1634 (2001)].
\end{abstract}

\vspace{12pt} 

\begin{multicols}{2}

The discrete nature of the energy spectrum in semiconductor quantum dots (QDs) 
is supposed to lead to a strong suppression of spin relaxation \cite{Gm} where
 promising applications for new spin dependent electronic devices have been 
predicted\cite{P,LV}. 
The current interest in manipulating semiconductor spins for applications 
is based on the ability to control and maintain spin coherence over 
practical length and time scales.
Optical pumping experiments have indeed given good indications of a
slowing down of the carrier spin relaxation processes in QD compared to 
bulk or quantum well structures.\cite{Ul,Hv} 
Recently, a detailed time-resolved investigation of the
intrinsic spin dynamics in $\rm InAs/GaAs$ QDs under strictly resonant 
excitation has been reported by Paillard et al.\cite{Pai} 
They demonstrated that at low temperature the carrier spins are totally 
frozen on the scale of the exciton lifetime. 
A rapid temporal decay of the linear polarization degree is, however, 
found above $30$K.

As a possible intrinsic mechanism for temporal decay of the linear 
polarization, we propose here the second-order quasielastic
interaction between QD carriers and LO phonons as sketched in Fig.\ 1(b). 
A similar mechanism was suggested by Uskov et al.\cite{Us} to explain the 
temperature dependence of a broadening of the zero-phonon line.
We will show that such relaxation processes are not suppressed (strongly) 
by the ``phonon bottleneck'' and can lead to a decay time of the 
linear polarization of some tens of picoseconds at $\simeq 100$K.

Let us first recall that self-organized QDs are usually strained and have 
an asymmetrical shape with a height smaller than the base size. 
The upper valence band in such QDs with the zinc--blende lattice is split 
into a heavy-hole band with the angular momentum projections 
$j_{h,z} = \pm 3/2$ and a light-hole band with  $j_{h,z} = \pm 1/2$ 
at the center of the Brillouin--zone (here the growth direction $z$ is 
chosen as the quantization axis).  
The conduction band  is $s$--like, with two spin states $s_{e,z} = \pm 1/2$.
The heavy-hole exciton quartet is characterized
by the total angular momentum projections $J_{z} = \pm 1$, $\pm 2$. 
The radiative states  $J_{z} = \pm 1$
and nonradiative ones,  $J_{z} = \pm 2$, are split by the  electron-hole 
exchange interaction (so--called singlet--triplet splitting 
$\Delta_{st}$).\cite{BP,Am,TE}
In the case of an asymmetric confinement potential in the plane of QDs 
the symmetry of the system is lowered and the exchange interaction is no 
longer isotropic.\cite{Gm,Iv} 
As a result,  both doublets are split into singlets, as is shown 
schematically in  Fig.\ 1.
The radiative doublet $|\pm 1\rangle$ is split by an anisotropic exchange
into the states labeled 
$|X\rangle = (|1 \rangle + \imath |-1 \rangle)/\sqrt{2}$
and  $|Y\rangle = (|1 \rangle - \imath |-1 \rangle)/\sqrt{2}$, 
giving rise to optical 
transitions which are linearly polarized along the $[110]$ and  
$[1\bar{1}0]$ directions, respectively.\cite{Iv1,BK}
Continuous wave single dot spectroscopy experiments have clearly evidenced
these two linearly polarized lines in self-organized 
$\rm InGaAs$ QDs. \cite{BK}
An analogous splitting of the exciton states has been observed in studies of
single QD's formed at $\rm GaAs/AlGaAs$ interfaces. \cite{Gm1} The observation
of the ``optical orientation-optical alignment'' or ``alignment-orientation''
transformation in a magnetic field makes it possible to 
determine the magnitude of the splitting without resolving the 
fine--structure spectrally.\cite{Pai,Iv1,Iv2} 
The anisotropic exchange splitting originates from the elongation of the 
QDs. \cite{Gm,Iv1,H}
The calculated and measured magnitudes of this splitting reach some tens
or even hundreds of $\mu$eV.\cite{Pai,Iv1,BK,Gm1} 

\vfill
\begin{figure}[b]
\centerline{\psfig{file=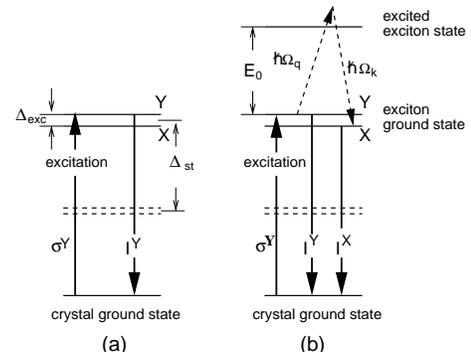,width=6cm}}

\vspace{6pt}
\caption{Schematic diagram showing (a) the sublevels of the exciton ground 
state and 
(b) the second-order phonon scattering process within the radiative doublet 
via the first excited state. Optically-inactive states are shown by dashed 
lines.}
\label{f1}
\end{figure}

In the case of (pulsed) resonant excitation which is linearly polarized, say, 
along the  $Y$--direction ($\sigma^{Y}$- excitation), an exciton in the  
$Y$--sublevel is created. If there is no relaxation to the $X$--sublevel,
the following emission occurs from the same $Y$- sublevel, 
as is shown in Fig.\ 1(a), and
the linear polarization degree of the luminescence
$P_{lin} = (I^{Y} - I^{X})/(I^{X} + I^{Y})$ (where $I^{X}$ and $I^{Y}$ 
denote the $X$ and $Y$ linearly polarized luminescence components) remains 
constant within the exciton lifetime, $P_{lin} = 1$. 
The possible relaxation process is shown schematically in Fig.\ 1(b). 
This is the second-order exciton-phonon scattering process in which one 
phonon  is absorbed and another one is emitted. 
The scattering events occur via the excited states of the exciton 
(with negligible exchange splitting) which can couple to both $X$-- and 
$Y$--exciton states. These processes lead to the exciton transitions 
between the ground $Y$-- and  $X$-- sublevels and result in the temporal 
dependence of $P_{lin}$. 
The (complete) rate equations governing the populations $f_{Y}$ and  
$f_{X}$ of the $Y$- and  $X$- sublevels read
\begin{eqnarray}
\frac{d f_{Y}}{d t} &=& - \frac{f_{Y}}{\tau_{rad}}  - \frac{f_{Y}}{\tau_{sc}} 
 + \frac{f_{X}}{\tau_{sc}}\;, \nonumber\\
\frac{d f_{X}}{d t} &=& - \frac{f_{X}}{\tau_{rad}}  - \frac{f_{X}}{\tau_{sc}} 
 + \frac{f_{Y}}{\tau_{sc}},
\label{R}
\end{eqnarray}
where $\tau_{sc}$ is the scattering time for an exciton relaxing from the  
$Y$($X$)--state to the  $X$($Y$)--state and  $\tau_{rad}$ is the radiative 
recombination time.
For the relevant initial conditions $f_{Y}(0) = f_{0}$ and  
$f_{X}(0) = 0$, Eqs.\ (\ref{R}) lead to an exponential decay of the
polarization degree
$P_{lin} = (f_{Y}-f_{X})/(f_{Y}+f_{X})=\exp(-t/\tau_{pol})$ with 
$\tau_{pol} = \tau_{sc}/2$.

As was noted above, in experiments by Paillard et al.\cite{Pai} no decay
of $P_{lin}$ is observed on the
exciton lifetime scale at $T < 30$K, but the decay time of $P_{lin}$ 
drops from $\sim 3.5$ns at $40$K down to $\sim 50$ ps at  $80$K.
In order to explain these findings we consider the contribution of 
LO phonons to the second-order scattering processes above.
In this case a strong (nonlinear) temperature dependence of a relaxation time 
$\tau_{sc}$ can be expected in the temperature range $k_BT$ smaller than the 
LO phonon energy $\hbar\Omega_{0}$.

Our calculation of the polarization decay time is based on the Fermi
golden rule for second order processes. To simplify matters,
we restrict ourselves to a certain range of parameters where the
following conditions hold:\\ 
(i) We assume that the phonon energy $\hbar \Omega_{0}$ is in the vicinity
of the excitation energy $E_0$ from the exciton ground 
state doublet (labelled ``0'') to the first excited  
 state (labelled ``1'') so that contributions of higher excited  
states can be neglected in the first approximation.
As a result, we have
\begin{eqnarray}
  \frac{1}{\tau_{sc}} = \frac{2 \pi}{\hbar} \;\sum_{\vec{q}, \vec{k}}  \;
  \Bigl|\frac{M_{\vec{q}}^{01} \; M_{\vec{k}}^{10}}
         {E_{0} + \hbar\Omega_{\vec{k}} + \imath\Gamma/2}
   + \frac{M_{\vec{k}}^{01}\;M_{\vec{q}}^{10}}
       {E_{0} -\hbar\Omega_{\vec{q}} + \imath\Gamma/2} \Bigr|^{2} 
       \nonumber\\
   \times N_{\vec{q}}\;(N_{\vec{k}} + 1)\;\delta(\Delta_{exc} - 
   \hbar \Omega_{\vec{k}} + \hbar \Omega_{\vec{q}}),
\label{W}
\end{eqnarray}
where $\vec{k}, \Omega_{\vec{k}}, \Gamma$ and 
$N_{\vec{k}} = 1/(e^{\hbar \Omega_{\vec{k}}/k_B T} - 1)$ are the phonon 
wave vector, frequency,  linewidth, and thermal distribution function, 
respectively.\\
(ii) To evaluate the exciton--phonon matrix elements  $M_{\vec{q}}^{01}$, 
we restrict ourselves to the Fr\"ohlich interaction\cite{Fr} of bulk LO 
phonons 
with carriers in strongly confined flat QDs (strained in the growth 
direction). For a strong confinement, the exciton states are
reasonably well approximated by noninteracting electron--hole pair states. 
In addition, the electron mass is much smaller than the heavy--hole mass 
and the first excited exciton state is mainly composed of a ground state 
electron and an excited hole. As a result, the electron does not couple 
to phonons because of the orthogonality of the initial and final hole states.
Thus, $M_{\vec{q}}^{01}$ reduces to the matrix element constructed 
with the hole envelope wave functions $\Psi^h_{0}(\vec{r})$  
and  $\Psi^h_{1}(\vec{r})$,
\begin{equation}
  M_{\vec{q}}^{01} = \imath \; 
    \frac{(2 \pi e^2 \hbar \Omega_{0})^{1/2}}{q \sqrt {V \kappa}}\;
    \langle\Psi^h_{0}(\vec{r}) |e^{\imath\;\vec{q}\; \vec{r}}| 
           \Psi^h_{1}(\vec{r})\rangle,
\label{ME}
\end{equation}
where $q = |\vec{q}|$, $V$ is the normalization volume, and
$1/\kappa = 1/\varepsilon_{\infty} - 1/\varepsilon_{0}$
($\varepsilon_{0}$ and $\varepsilon_{\infty}$ denote the static and
high--frequency limits of the dielectric function).\\
(iii) Equation\ (\ref{ME}) is mostly affected by the oscillatory behavior
of the exponential function rather than by the detailed shape of the
wave functions. 
Therefore, we may furthermore neglect a weak lateral anisotropy of the QD and, 
for simplicity, consider a parabolic potential of the form
$\sim \alpha_{\parallel}^4 (x^2 + y^2) + \alpha_{z}^4 z^2$.
The potential strengths $\alpha_{\parallel}$ and  $\alpha_{z}$ are inversely 
proportional to the base length $L_{\parallel}$ and a height $L_{z}$ of the 
QD, respectively. In this case the hole envelope  wave functions are given by
\begin{eqnarray}
\Psi^h_{0}(\vec{r}) &=& 
      \frac{\alpha_{\parallel} \alpha_{z}^{1/2}} {\pi^{3/4}}\;
      \exp\bigl\{- \frac{\alpha_{\parallel}^{2}}{2} (x^2 + y^2)  
                 - \frac{\alpha_{z}^2}{2} z^2\bigr\}, \nonumber\\
\Psi^h_{1}(\vec{r}) &=& 
     \sqrt{2} \alpha_{\parallel} \; x \; \Psi^h_{0}(\vec{r}).
\label{wf}
\end{eqnarray}
In addition, we have neglected the heavy--hole--light--hole
mixing because in the case of strained flat QDs the energy separation between 
the first heavy--hole and light--hole levels is large compared to the 
lateral confinement energies.\\
(iv) It follows from Eqs.\ (\ref{ME},\ref{wf}) that only LO--phonons 
with small wave numbers 
$q \lesssim \alpha_{\parallel}\sim L_{\parallel}^{-1}$ contribute 
substantially to the scattering rate Eq.\ (\ref{W}). 
Therefore, the phonon dispersion can be ignored in the Bose functions, 
$N_{\vec{k}}, N_{\vec{q}} \rightarrow N_{0}$. 
In the delta--function, however, the dispersion is essential 
and it will be approximated by an isotropic relation, 
$\hbar \Omega_{\vec{k}} = \hbar \Omega_{0} - v k^{2}$.
Below, we restrict our consideration to energy window
%the relatively large energy differences
$0.5\hbar\Omega_0\geq |E_{0} - \hbar \Omega_{0}| 
 \geq 0.05 \hbar\Omega_{0} \approx 1.5$meV 
($\hbar \Omega_{0} \simeq 30$meV). 
In this case $|E_{0} - \hbar \Omega_{0}|$ is 
larger than the exchange splitting $\Delta_{exc}$, 
the effective width of the phonon band $v \alpha_{\parallel}^2$, 
as well as the phonon line width $\Gamma$.\cite{estim} 
Therefore,  these contributions can be neglected in the energy 
denominators of Eq.\ (\ref{W}).

\begin{figure}[t]
\centerline{\psfig{file=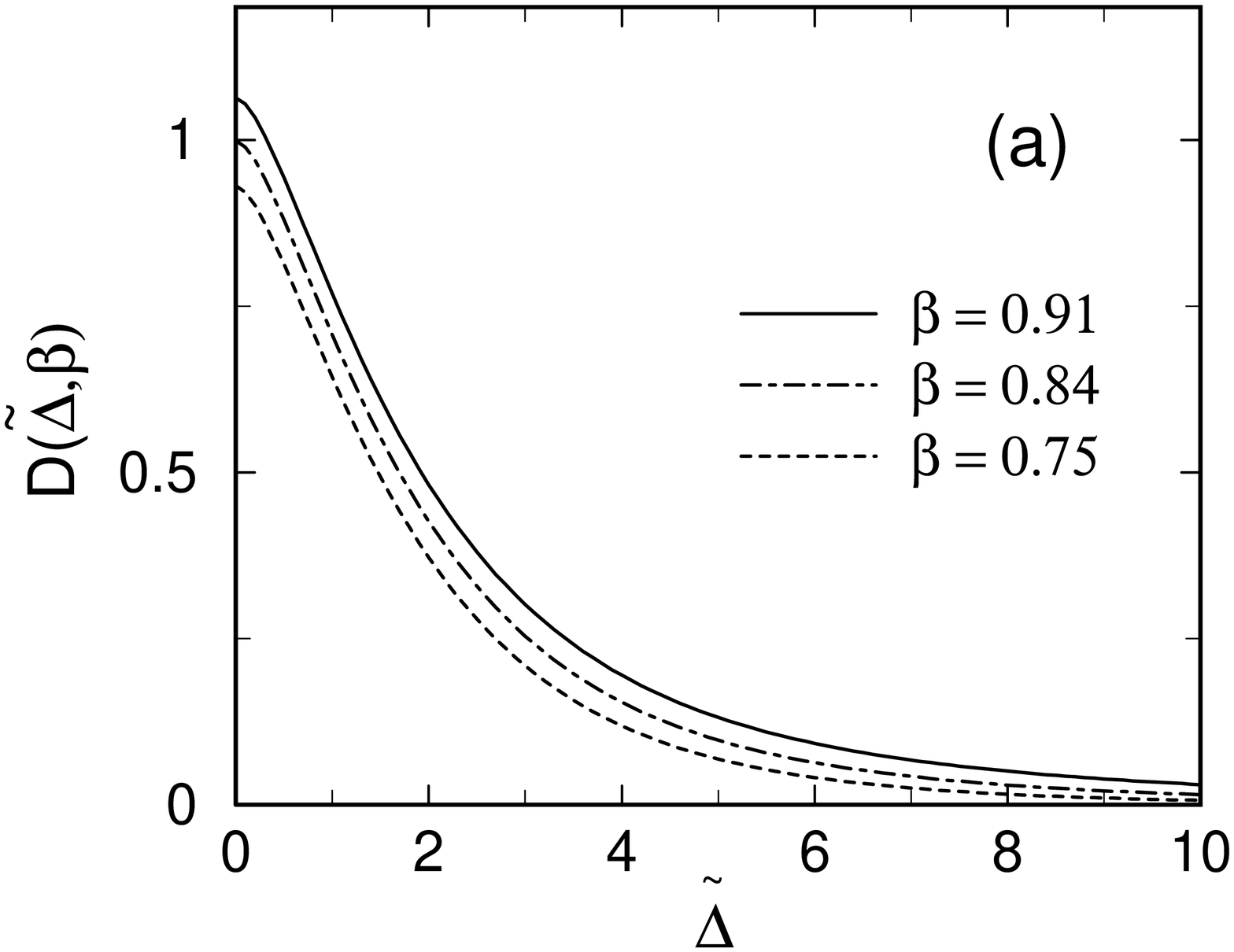,width=6.5cm,angle=270}\quad}
\centerline{\psfig{file=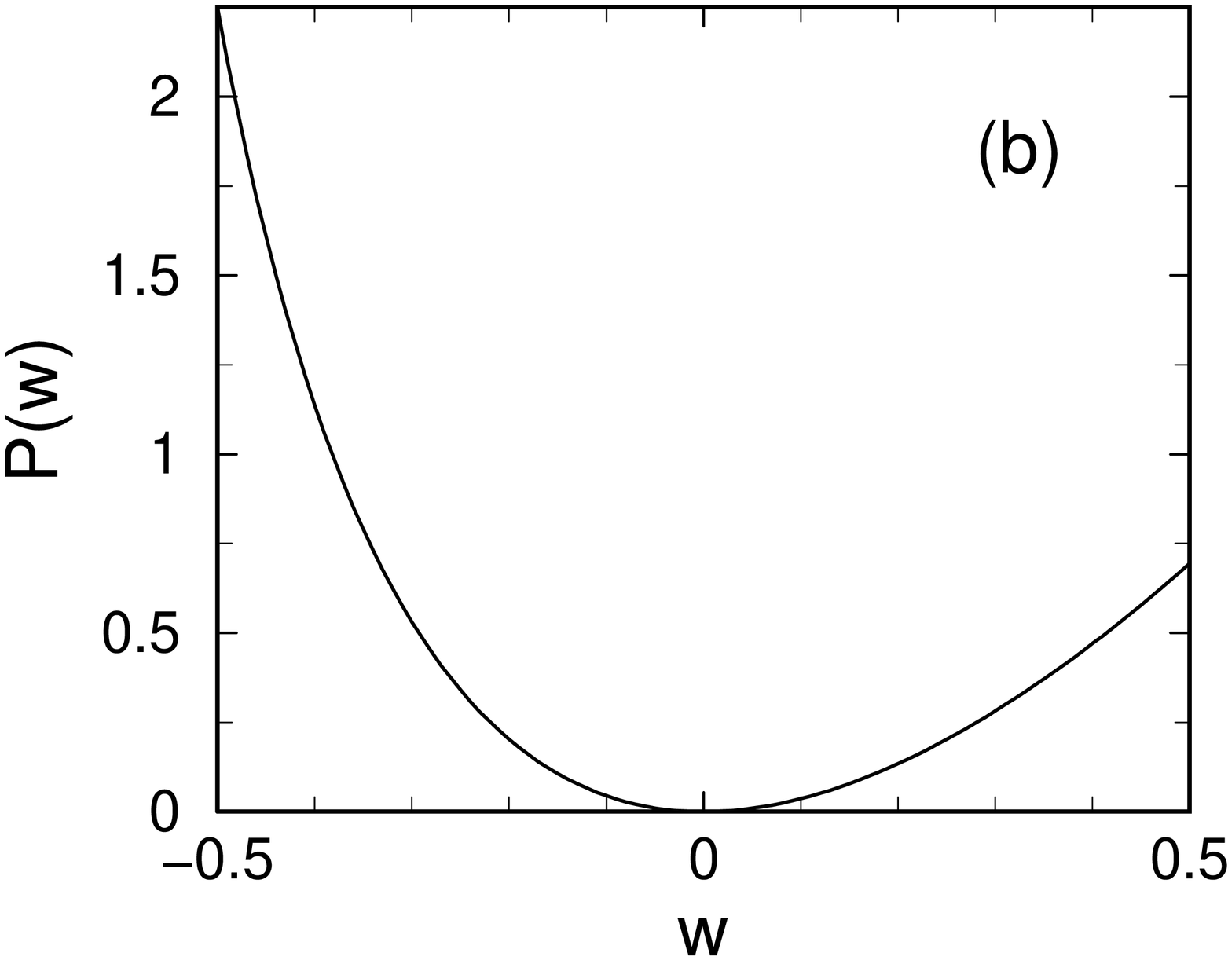,width=6.3cm,angle=270}}

\caption{Phonon (a) and level detuning (b) functions.}
\label{f2}
\end{figure}

After some algebra, we obtain our main result
\begin{eqnarray}
\tau_{pol} &=& \tau_0\;\frac{1}{N_0(N_0+1)}\;\frac{P(w)}{D(\tilde{\Delta}, 
              \beta)}, \label{Wi}\\
\tau_0 &=& \sqrt{2} (2)^{6} \pi \frac{\kappa^2 \hbar v}{e^4},\label{tau0}\\
P(w)  &=& \Bigl[\frac{w (w+2)}{(w+1)}\Bigr]^2,\label{P}\\
D(\tilde{\Delta}, \beta) &=& e^{-\tilde{\Delta}}\;
    \int_{0}^{\infty}d x x^2 e^{-x^2} F(\frac{x^2 \beta}{2})\nonumber\\
    &~& \times F(\beta(\tilde{\Delta}+ \frac{x^2}{2})) 
     \sqrt{\tilde{\Delta}+ \frac{x^2}{2}},\label{Delta}\\
F(z) &=& \sqrt{\pi/z} \;\mbox{erfi}(\sqrt{z})\;
         (1 + 1/2 z) - e^{\sqrt{z}}/2 \sqrt{z},\label{Wi1}
\end{eqnarray}
where $\tau_0$ is the characteristic decay time of the QD which depends on the
electron--phonon coupling strength $\kappa$ and the phonon dispersion parameter
$v$.
The parameter $w = (E_{0} - \hbar \Omega_{0})/\hbar \Omega_{0}$ defines 
the ``phonon detuning'' between
the interlevel spacing in the QD and the LO phonon energy, whereas
$\tilde{\Delta} = \Delta_{exc}/(2 v \alpha_{\parallel}^2)$ defines the 
``mismatch'' between the
radiative level separation and the effective width of the phonon 
band\cite{ET,VFB}.
The parameter $\beta = 1 - (L_{z}/L_{\parallel})^2$ measures the anisotropy 
of the QD (for flat QDs under consideration $\beta>0$ and close to one) 
and $\mbox{erfi}(z) = \mbox{erf}(\imath z)/\imath$ is the imaginary 
error function\cite{Mat}.

From the structure of Eq. (\ref{Wi}) we can draw qualitative conclusions
about the influence of temperature, level and phonon--detunings, and size and 
shape of the QD on the polarization decay time. 

\begin{figure}[t] 
\centerline{\psfig{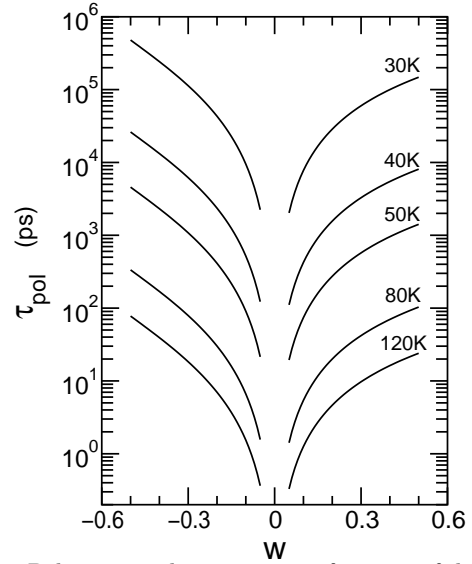}}

\caption{Polarization decay time as a function of
the phonon detuning parameter for ${\tilde\Delta}=5$ and
$\beta=0.75$.}
\label{f3}
\end{figure}

The temperature dependence is determined 
by the phonon number $N_{0}$ and $\tau_{pol}$ decreases 
exponentially with temperature if $k_{B}T \ll \hbar \Omega_{0}$.

The level detuning function  for flat QDs with 
the anisotropy parameter $\beta =0.91 (L_{z}/L_{\parallel} = 0.3)$, 
$\beta =  0.84 (L_{z}/L_{\parallel} = 0.4)$, and $\beta = 0.75 
(L_{z}/L_{\parallel} = 0.5)$ is shown in Fig.\ 2(a).
$D(\tilde{\Delta}, \beta)$ decreases monotonically  with  $\tilde{\Delta}$ and 
has an exponential limit for $\tilde{\Delta} \gg 1$.
Note that $D(\tilde{\Delta}, \beta)$ enters in Eq. (\ref{Wi}) due to
the condition of the energy conservation during the scattering process 
(cfm. $\delta$-function in Eq. (\ref{W})).
Furthermore, the level splitting $\Delta_{exc}$  plays  the role  of the 
energy difference $(\hbar \Omega_{0} - E_{0})$ when compared with an 
inelastic first-order scattering process.\cite{IS} 
As it is well known, such processes are strongly suppressed by 
the phonon bottleneck effect.
The second--order processes under consideration are, however,  more efficient.
The point is that an anisotropic exchange splitting 
$\Delta_{exc}\sim 0.1$meV is by itself relatively small\cite{Pai,BK},
while some specific conditions are required to obtain similar magnitudes 
for the energy difference $(\hbar \Omega_{0} - E_{0})$. 
In addition, the parameter 
$\tilde{\Delta}=(\Delta_{exc}/ v \alpha_{\parallel}^2)$ 
is not expected to change very strongly with the lateral size of the QD  due 
to the size 
dependence of the exchange splitting $\Delta_{exc}$.\cite{Iv1}$^{,}$\cite{Tg} 
Thus,  $\tilde{\Delta}$ is determined mostly by the lateral anisotropy of 
the QDs and can be
of the order of unity (using for $\Delta_{exc}$ and 
$v \alpha_{\parallel}^{2}$ the values
$\Delta_{exc}\sim 0.1$meV and $v \alpha_{\parallel}^{2}\sim 0.01$meV, 
noted above, one obtains that $\tilde{\Delta} \sim 5$). 
It can be seen in Fig.\ 2(a) also that the level detuning function 
$D(\tilde{\Delta}, \beta)$ is modified
by changing the ratio $L_{z}/L_{\parallel}$.  For instance, it increases
by a factor of $\sim 1.3$ at $\beta \sim 0.91$ compared to 
$\beta \sim 0.75$  in the case of $\tilde{\Delta} = 2$,  
by a factor of $\sim 1.9$ at $\tilde{\Delta} = 5$, and by a factor 
of $\sim 2.6$ at $\tilde{\Delta} = 7$. 

The size dependence of polarization relaxation is determined by the
phonon detuning function $P(w)$ which is shown in Fig.\ 2(b).
The nonexponential dependence of $P(w)$ on $w$  results in a relatively weak
dependence  of the polarization decay on the phonon detuning. 
In addition, even at relatively large detunings, e.g.,
$w \simeq 0.5$  or $w \simeq -0.3$, $P(w)$ is of the order of unity. 
The relaxation occurs for both positive ($E_{0} > \hbar \Omega_{0}$) 
and negative ($E_{0} < \hbar \Omega_{0}$) detunings but there is some 
asymmetry.

Below we give some numerical estimates of the polarization decay time
in  $\rm InAs$ QDs which can be considered as typical for III--V compounds.
Material parameters used are: $\hbar \Omega_{0} \simeq 30.2$meV,  
$\epsilon_0\simeq 15.15$, $\epsilon_\infty\simeq 12.25$,\cite{LB} and 
$v \simeq 0.1$(nm)$^2$meV \cite{ET} which implies  $\tau_0 \simeq 0.07$ps.
For $T \ll 350$K  the number of phonons $N_{0}$ is very small so that 
$\tau_{pol}$ decreases exponentially with temperature, 
e.g., by a factor of $\simeq 80$ between $T=40$K and $T=80$K
which is in agreement with experiment.\cite{Pai}
Because the numerical values of phonon and level detuning, 
as well as the QD anisotropy, are usually not known exactly from experiments, 
$w$,  $\tilde{\Delta}$ and $\beta$ are used as adjustable parameters. 
As an illustration, the  decay time of the polarization degree 
is shown in Fig.\ 3  as a function of the phonon detuning parameter 
$w$ at different temperatures. The decay time
$\tau_{pol}$ drops from a few hundred (units) of nanoseconds at $T = 30$K down 
to a few tens (and even units) of picoseconds at $T = 120$K, 
depending on the phonon detuning.
For $w = - 0.15$ (i.e. $E_{0}-\hbar\Omega_0 \simeq - 4.5$meV), 
$\tilde{\Delta} = 7$ and $\beta = 0.75$
which seem to be plausible for the experiments by Paillard et al.,\cite{Pai} 
the calculated decay times $\tau_{pol}\simeq 3.5$ns ($40$K) and 
$\tau_{pol}\simeq 44$ps (80K) closely agree with the experimental results.  

In conclusion, the second-order quasielastic interaction between charge 
carriers and LO phonons in strongly confined asymmetric QDs
is identified as an intrinsic mechanism of the temporal decay of the 
linear polarization degree of the luminescence.
Despite of the apparent obstruction by the phonon bottleneck and the level 
detuning effects, the proposed mechanism leads to decay times of 
the order of a few tens of picoseconds (or even smaller) at 
temperature  $T \simeq 100$K which are in agreement with experiments. 
The relaxation processes are more efficient in flat QDs with a weak 
lateral anisotropy.
Equation\ (\ref{Wi}) may be useful for the optimation of 
QD--structures for application in spintronic devices.

%\vspace{12pt}
This work was supported by the Center for Functional Nanostructures (CFN) 
of the Deutsche Forschungsgemeinschaft (DFG) within project A2.

\end{multicols}


\begin{references}
\bibitem{TE}
  Generally speaking, the nonradiative doublet is split also by the exchange 
  interaction into two, close lying, optically inactive singlets.
\bibitem{estim}
  For a typical QD the exchange splitting is 
  $\Delta_{exc} \sim 0.1$meV (Refs.\cite{Pai,BK}),
  the effective width of the phonon band is 
  $v \alpha_{\parallel}^2 \sim 0.01$meV (Ref.\cite{ET}), 
  and the phonon line width is $\Gamma\sim 0.1$meV (Ref.\cite{VFB}).
\bibitem{ET}
  We estimate the LO--phonon dispersion constant $v$ as 
  $v \simeq \Delta_{ph} (a/\pi)^2$, where $\Delta_{ph}$ is a width of the 
  phonon band and $a$ is the lattice constant. For typical values of 
  $\Delta_{ph} \simeq 5$meV and $a \simeq 0.5$nm we have 
  $v \sim 0.1$meV nm$^2$. The lateral size of a typical QD is usually
  more than an order of magnitude larger than $a$, so that the effective 
  width of the phonon band is typically $v \alpha_{\parallel}^2 \sim 0.01$meV.
\end{references}
\end{document}